\documentstyle[amsmath,amssymb,epsfig,12pt]{article}

\begin{document}
\setlength{\parskip}{0.45cm}
\setlength{\baselineskip}{0.75cm}
%
%
%
\begin{titlepage}
\setlength{\parskip}{0.25cm}
\setlength{\baselineskip}{0.25cm}
\begin{flushright}
DO-TH 2005/01\\
\vspace{0.2cm}
January 2005
\end{flushright}
\vspace{1.0cm}
\begin{center}
\Large
{\bf Radiatively Generated Isospin Violations}
\\\Large{\bf in the Nucleon}
\vspace{1.5cm}

\large
M.\ Gl\"uck, P.\ Jimenez--Delgado, E.\ Reya\\
\vspace{1.0cm}

\normalsize
{\it Universit\"{a}t Dortmund, Institut f\"{u}r Physik,}\\
{\it D-44221 Dortmund, Germany} \\
\vspace{0.5cm}

\vspace{1.5cm}
\end{center}

\begin{abstract}
\noindent Isospin violating valence and sea distributions are evaluated
due to QED leading ${\cal{O}}{(\alpha)}$ corrections to the standard QCD
evolution equations.  Unique perturbative predictions are obtained within
the radiative parton model, and confirm earlier results.  Nonperturbative contributions have been estimated and depend on a single free parameter chosen to be the current
quark mass.  The relevance of our predictions for extracting $\sin^2 \theta_W$
from DIS $\nu(\bar{\nu})N$ data (`NuTeV anomaly') is discussed as well. 
\end{abstract}
\end{titlepage}


The NuTeV collaboration recently reported \cite{ref1} a measurement of
the Weinberg angle $s_W^2\equiv \sin^2\theta_W$ which is approximately
three standard deviations above those presented in \cite{ref2} for the 
world average of other electroweak measurements.  Possible sources for
this discrepancy (see, for example, \cite{ref3,ref4,ref5,ref6,ref7})
include, among other things, an isospin violating contribution 
$\delta R_I^-$ to the relation \cite{ref8} $R_{PW}^-=\frac{1}{2}-s_W^2$,
\begin{equation}
R^-\equiv \frac{\sigma_{NC}^{\nu N} - \sigma_{NC}^{\bar{\nu}N}}
   {\sigma_{CC}^{\nu N} - \sigma_{CC}^{\bar{\nu}N}}
= \frac{1}{2} -s_W^2 +\delta R_I^-\, .
\end{equation}
This contribution is given, for $N=\frac{1}{2}(p+n)$, by 
\cite{ref3,ref5,ref7}
\begin{equation}
\delta R_I^-=\left( \frac{1}{2}-\frac{7}{6}\,\,s_W^2\right)\,\,
  \frac{\delta U_v-\delta D_v}{U_v+D_v}
\end{equation}
where the second moments $\delta Q_v(Q^2)$ and $Q_v(Q^2)$ of the valence
distributions $(\delta)q_v(x,Q^2)$ are $Q_v(Q^2)=\int_0^1 x\, q_v(x,Q^2)dx$,\,\,
$q_v=q-\bar{q}$, and $\delta Q_v(Q^2)=\int_0^1x\, \delta q_v(x,Q^2)dx$ with
\begin{eqnarray}
\delta u_v(x,Q^2) & = & u_v^p(x,Q^2)-d_v^n(x,Q^2)\nonumber\\
\delta d_v(x,Q^2) & = & d_v^p(x,Q^2)-u_v^n(x,Q^2)\, .
\end{eqnarray}
NLO QCD corrections do not significantly \cite{ref4,ref6,ref7} contribute
to (1).  Nonperturbative calculations \cite{ref9,ref10,ref11} of $\delta q_v$
in (3) were found \cite{ref5} to possibly reduce the discrepancy between the
neutrino and electroweak measurements of $s_W^2$ by up to 40\%.  A somewhat
different calculation \cite{ref12}, based on QED contributions to 
$\delta q_v(x,Q^2)$, resulted in a comparable reduction of this discrepancy.

We shall approach the issue of the isospin violations within the
framework of the radiative parton model \cite{ref13} and obtain
predictions for $\delta q_v$ which depend on a single free parameter
required by the nonperturbative contribution and chosen to be, as in
\cite{ref12}, the current quark mass.  Although our method differs from
those in \cite{ref5}, the results turn out to be similar. On the other hand, our starting position is only a little different to \cite{ref12}, the results, as anticipated in that work, turn out to be quite similar.

Following \cite{ref12} we shall evaluate the modifications of the 
parton distributions due to QED radiative effects. In contrast to
\cite{ref12,ref14} we shall calculate these effects only to 
{\underline{leading}} order in $\alpha$, which is sufficient for our
purpose here and furthermore simplifies the calculations considerably.
The inclusion of QED ${\cal{O}}(\alpha)$ corrections modifies the QCD
evolution equation for charged parton distributions by an additional
term which, in an obvious symbolic notation, reads \cite{ref15,ref16}
\begin{equation}
\dot{q}(x,Q^2) = \frac{\alpha_s}{2\pi}
 \left[ P_{qq}*q + P_{qg}*g\right] + \frac{\alpha}{2\pi}
  \,\, e_q^2P_{qq}^{\gamma}*q
\end{equation}
with $P_{qq}^{\gamma}(z)=\left(\frac{1+z^2}{1-z}\right)_+$.
Notice that the addition {\cite{ref12,ref14} of a further term 
$(\alpha/2\pi)\,e_q^2\, P_{q\gamma}*\gamma$ to the r.h.s.\ of eq.~(4)
would actually amount to a subleading ${\cal{O}}(\alpha^2)$ contribution
since the photon distribution $\gamma(x,Q^2)$ of the nucleon is of
${\cal{O}}(\alpha)$ \cite{ref17,ref18,ref19,ref20,ref21,ref22}.
The standard QCD evolution  equation for the gluon distribution remains
unaltered at leading ${\cal{O}}(\alpha)$.  The perturbative $(p)$
${\cal{O}}(\alpha)$ modifications of the valence quark distributions
in (3), as implied by (4), are now given by
\begin{eqnarray}
\delta_p u_v(x,Q^2) & = & \frac{\alpha}{2\pi} 
      \int_{Q_0^2}^{Q^2}d\ln q^2 
       \int_x^1\frac{dy}{y}\,\, 
         P\left(\frac{x}{y}\right) u_v(y,\, q^2)\nonumber\\
\delta_p d_v(x,Q^2) & = & -\frac{\alpha}{2\pi}
      \int_{Q_0^2}^{Q^2}d\ln q^2 
       \int_x^1\frac{dy}{y}\,\, 
         P\left(\frac{x}{y}\right) d_v(y,\, q^2)
\end{eqnarray}
with $P(z)=(e_u^2-e_d^2)P_{qq}^{\gamma}(z)$ and 
$Q_0^2\equiv \mu_{\rm LO}^2 = 0.26$ GeV$^2$, $u_v(x,q^2)$, $d_v(x,Q^2)$
are taken from the dynamical (radiative) parton model \cite{ref13}. Notice that this treatment of the perturbative components differs from that in \cite{ref12}.

For the nonperturbative ($np$) modifications we estimate, following eq.\,(12) in \cite{ref12}, 
\begin{eqnarray}
\delta_{np} u_v(x) & = & \frac{\alpha}{2\pi} \,\,
      \ln \frac{Q_0^2}{\mu_0^2} 
       \int_x^1\frac{dy}{y}\,\, 
         P\left(\frac{x}{y}\right) u_v(y,\, Q_0^2)\nonumber\\
\delta_{np} d_v(x) & = & -\frac{\alpha}{2\pi}\,\,
      \ln \frac{Q_0^2}{\mu_0^2} 
       \int_x^1\frac{dy}{y}\,\, 
         P\left(\frac{x}{y}\right) d_v(y,\, Q_0^2)
\end{eqnarray}
where we take $\mu_0=m_q\simeq 10$ MeV, i.e.\ of the order of the 
current quark masses \cite{ref2}.  Here the parton distributions at
$Q^2\leq Q_0^2$ were taken to equal their values at the perturbative
input scale $Q^2=Q_0^2$, i.e.\ were `frozen´'.

In fig.~1 we show the total $\delta_p\, q_v(x,Q^2)+\delta_{np}\, q_v(x)$
and the purely perturbative $\delta_p\, q_v(x,Q^2)$ isospin violating
distributions at a typical scale of $Q^2=10$ GeV$^2$.  Our total isospin
violating distributions are quite similar to those in \cite{ref12}, as well as to those in \cite{ref5} which were obtained by rather different methods.  The fact that only our 
{\underline{total}} results are compatible with nonperturbative (bag)
model expectations \cite{ref5,ref9,ref10,ref11} is indicative for the 
necessity
of our nonperturbative estimates in (6) and that the rather marginal
perturbative contributions in (5) are not sufficient for a realistic
estimate of isospin violating effects.  It is, furthermore, particularly
interesting to note that our results depend, as mentioned above, only
on a {\underline{single}} free parameter, i.e.\ $\mu_0$ in (6).

Encouraged by this agreement we present in fig.~2 and 3 the corresponding
predictions for the isospin violating distribution $\delta\bar{u}$ and
$\delta\bar{d}$ of sea quarks as defined in (3) and obtained from (5)
and (6) by replacing $u_v$ and $d_v$ by $\bar{u}$ and $\bar{d}$,
respectively.  At $Q^2=10$ GeV$^2$ the perturbative contribution
$\delta_p\bar{q}(x,Q^2)$ in fig.~2 does not dominate the total result
$\delta_p\bar{q}(x,Q^2)+\delta_{np}\bar{q}(x,Q^2)$ which is obviously
in contrast to the predictions at higher scales as illustrated, for
example, at $Q^2=M_W^2$ in fig.~3.  Similar results are obtained for 
the LO CTEQ4 parton distributions \cite{ref23} where also valence--like
sea distributions are employed at an input scale $Q_0^2=0.49$ GeV$^2$,
i.e., $x\bar{q}(x,Q_0^2)\to 0$ as $x\to 0$.  Such predictions may be
tested by dedicated precision measurements of Drell--Yan and DIS
processes employing neutron (deuteron) targets as well.

As a possible immediate application of our predictions for the 
isospin violating valence distributions let us finally turn to the 
relation in (1).
Since the isospin violation generated by the QED ${\cal{O}}(\alpha)$
correction is such as to remove more momentum from up--quarks than
down--quarks, as is evident from fig.~1, it works in the right direction
to reduce the NuTeV anomaly \cite{ref1}, i.e.\ 
$\sin^2\theta_W = 0.2277\pm 0.0013\pm 0.0009$ as compared to the 
world average of other measurements \cite{ref2} 
$\sin^2\theta_W = 0.2228(4)$.  This effect is also slightly $Q^2$
dependent because of the perturbative contribution in (5).  At 
$Q^2 = 10$ GeV$^2$, appropriate for the NuTeV experiment, our total
distributions in fig.~1 imply $\delta U_v= -0.002226$ and 
$\delta D_v=0.000890$
which, together with $U_v+D_v=0.3648$ in (2), leads to a change in 
the measured value of $\sin^2\theta_W$  of
$\delta\sin^2\theta_W=\delta R_I^-=-0.0020$ according to (1).
Since the value of $\sin^2\theta_W$ from the NuTeV experiment, with
$\delta R_I^-\equiv 0$ in (1), is 0.0049 larger than the world average
of other measurements, our predicted isospin violation 
$\delta R_I^-$ reduces this discrepancy by about 40\%.  This reduction
may be slightly diminished if it is corrected for experimental
acceptance cuts \cite{ref3}.

To summarize, we evaluated the modification of parton distributions
due to QED leading ${\cal{O}}(\alpha)$ corrections to the standard
QCD evolution equations.  For the isospin \mbox{violating} valence 
$\delta q_v(x,Q^2)$ and sea $\delta\bar{q}(x,Q^2)$ distributions 
($q=u,\, d$)
unique perturbative predictions are obtained within the dynamical
(radiative) parton model.  The nonperturbative contributions to
$\delta q_v$ and $\delta\bar{q}$ have been estimated and depend on
a single free parameter chosen to be the current quark mass.  The
results for the valence distributions bear some similarity to 
nonperturbative bag--model expectations. Our total predictions for
$\delta q_v$ reduce significantly the discrepancy between the large
result for $\sin^2\theta_W$ as derived from deep inelastic 
$\nu(\bar{\nu})N$ data (`NuTeV anomaly') and the world average of 
other measurements.


\newpage
\begin{figure}
\epsfig{file=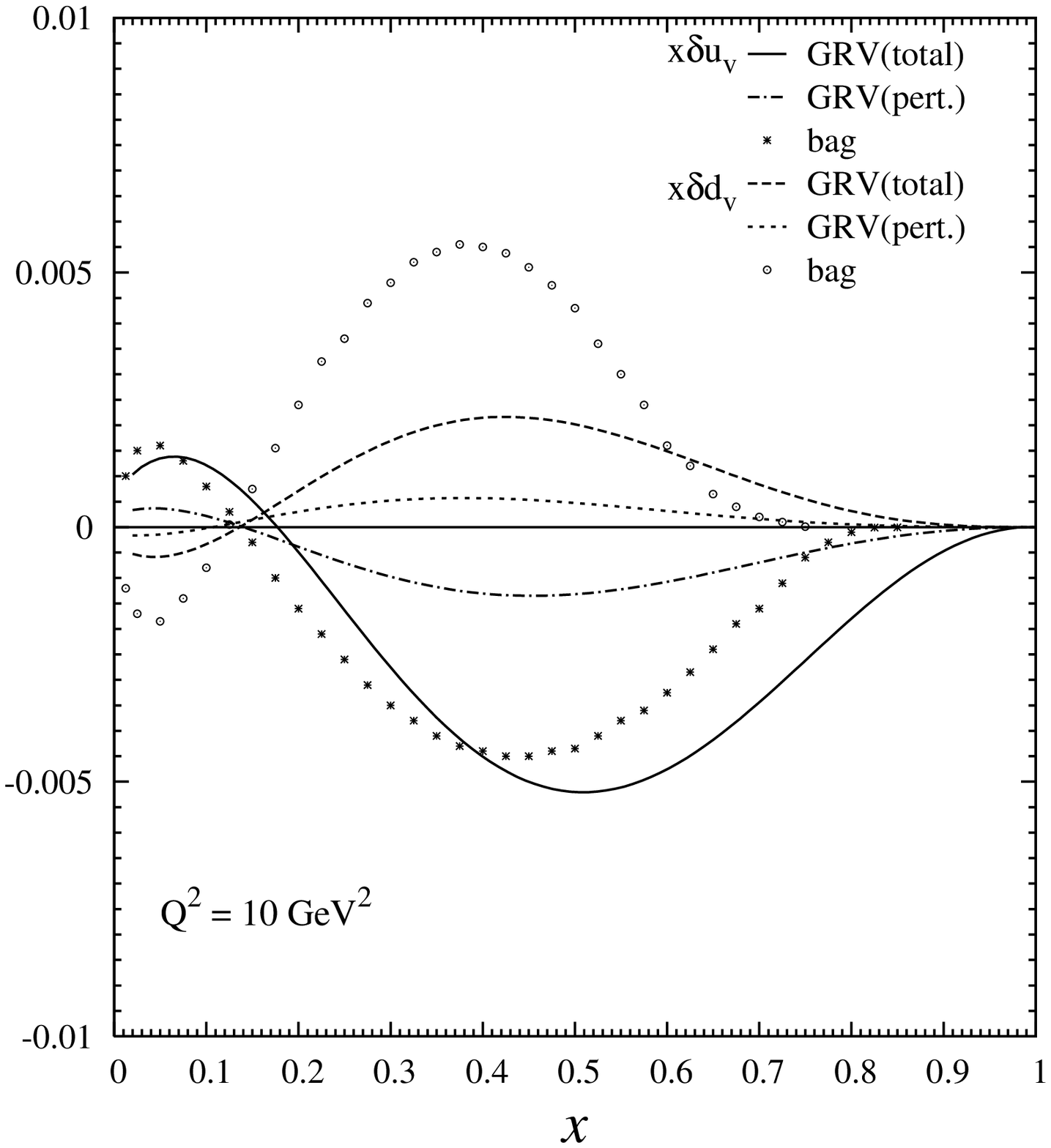,width=\textwidth}
\caption{The isospin violating `majority' $\delta u_v$
      and  `minority' $\delta d_v$ valence quark distributions at
      $Q^2=10$ GeV$^2$ as defined in (3).  The perturbative GRV(pert.) 
      predictions are calculated according to (5).  With the 
      non--perturbative contribution in (6) added, one obtains the 
      total estimates GRV(total).  The bag model estimates are taken
      from ref.~\cite{ref5}.}
\end{figure}

\newpage
\begin{figure}
\epsfig{file=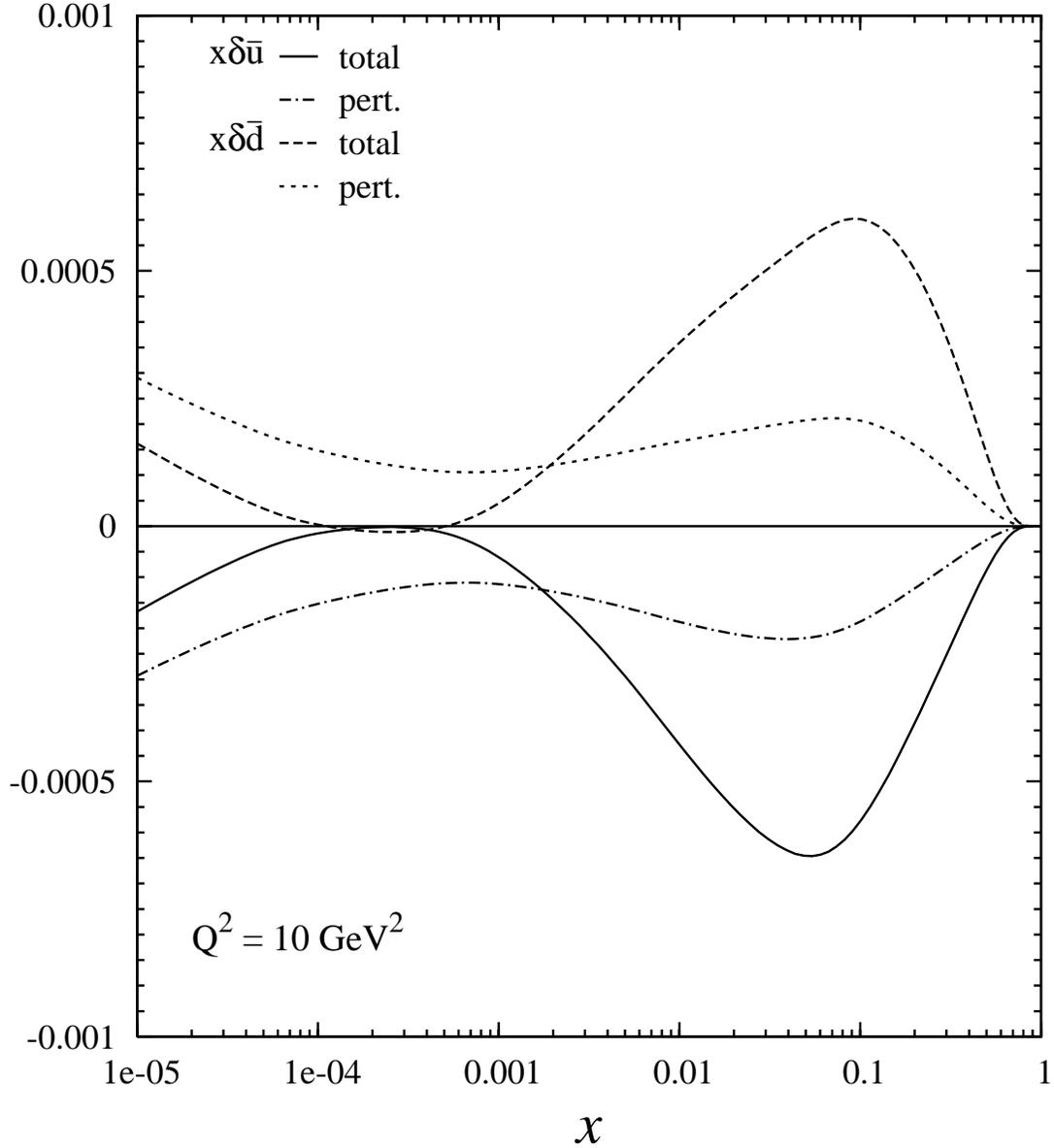,width=\textwidth}
\caption{The isospin violating sea distributions 
      $\delta\bar{u}$ and $\delta\bar{d}$ at $Q^2=10$ GeV$^2$ as
      defined in (3) with $u_v$, $d_v$ replaced by $\bar{u},\, 
      \bar{d}$.  The perturbative (pert.) predictions are calculated
      according to (5) with $u_v,\, d_v$ replaced by $\bar{u},\, \bar{d}$.
      The same replacement holds for (6) when calculating
      the non--perturbative contributions which have to be added to the 
      `pert.' predictions in order to obtain the total results. The LO GRV sea distributions \cite{ref13} are used throughout.}
\end{figure}

\newpage
\begin{figure}
\epsfig{file=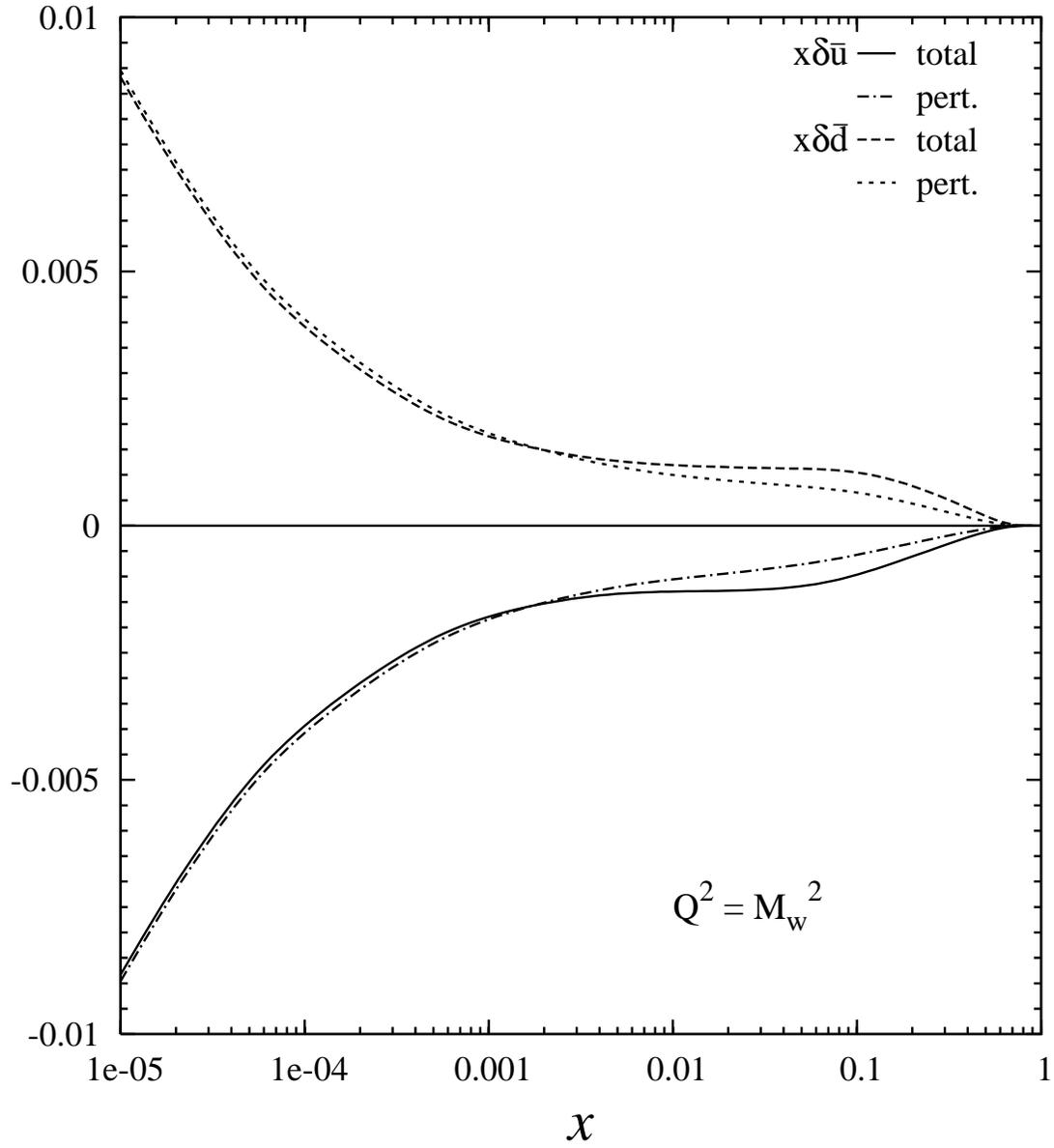,width=\textwidth}
\caption{As in fig.\ 2 but at $Q^2=M_W^2$.}
\end{figure}

\end{document}